\documentclass[varenna]{cimento}

\usepackage{graphicx}  
\title{Polarization catastrophe at low densities of polarons: 
from Cuprates to Metal-Ammonia Solutions}
\author{P. Qu\'emerais, S. Fratini}
\institute{quemerai@grenoble.cnrs.fr, 
fratini@grenoble.cnrs.fr\\
Laboratoire d'Etudes des Propri\'et\'es Electroniques des
  Solides, CNRS, Grenoble\\
BP 166, F-38042 Grenoble cedex 9, France}
\shortauthor{P. Qu\'emerais, S. Fratini}

\PACSes{\PACSit{00.00}{}
\PACSit{---.---}{\ldots}}
\begin{document}

\maketitle

\begin{abstract}
We review some results on the role played by dielectric 
polarons at the metal-insulator
transition in polarizable materials, taking into account the long-range 
nature of the Coulomb interactions. The occurrence of a polarization 
catastrophe is examined in a model describing a Wigner crystal of
polarons. The possible relevance of this scenario in the cuprates
and in metal-ammonia solutions is discussed. 
\end{abstract}

\section{Introduction}

Many mechanisms have been proposed to understand the microscopic
origin of the metal-insulator transitions
(MIT) occurring in condensed matter. Since the discovery 
of high-Tc superconductors in 1986,
the MIT occurring in the cuprates remains the focus of debates and new
theoretical studies. 
The main idea in the original work of M\"uller and Bednorz
 \cite{muller}, was to search conditions for strong
electron-phonon coupling. For this reason, they focused on
transition-metal oxides, a somehow heretic point of view at that time,
since such compounds 
belong to the category of insulators, or bad conductors.
The presence of antiferromagnetism  in the insulating phase made
physicists focus on Hubbard and related models  \cite{hubbard}, 
where a magnetically ordered insulating phase occurs 
due to short range
electronic correlations. However, 
the Hubbard  model is able to describe "bad metals"
(more precisely, metals with reduced electronic bands and with short
range electronic correlations), but, strictly speaking, not
insulators. The reason is that 
when a system of electrons is
insulating, long range Coulomb interactions are not screened and must
be incorporated in any model which properly describes such a state:
in other words, Hubbard's model is not self-consistent through the
MIT (away from half filling), as it
becomes  metallic already at very low
doping levels. 

The cuprates must be doped to finite densities of carriers to become
superconducting. One can sweep out this question by invoking impurities
or disorder (through Anderson localization), but up to now, there are
no conclusive studies 
able to understand this
particular and important point. Several studies have tried to
incorporate long range Coulomb effects in the theoretical description 
of the phenomenon, 
 but the basic mechanism of the
superconducting transition, as well as the origin of the necessary finite 
doping density, still remain unclear.
This fact is not restrained to the cuprates. In the bismuthates, where
high temperature superconductivity  ($T_c \approx 40K$) is observed
with no trace of magnetism (which rules out the Hubbard model) 
the same phenomenon also occurs: 
K$_x$Ba$_{1-x}$BiO$_3$ is insulating for $x<0.35$ and becomes
superconducting for $x>0.35$. 

Another important point is that the parent cuprates, as any insulating
oxides and more generally any iono-covalent insulating compounds,
belong to the category of  polar materials, as testified by the 
large differences between the static and high-frequency 
dielectric constants. For undoped La$_2$CuO$_4$  \cite{chen}, for example, 
it is  $\epsilon_s \approx 30$ in the
static limit, and approaches $\epsilon_{\infty} \approx 5$ at high frequency. 
This takes its origin from the presence of ionized atoms in their
structures, which makes the
difference with other covalent insulators or semiconductors such as
Si or GaAs, where the dielectric constant depends weakly on the
frequency. As is well-known since the work of Fr\"ohlich
 \cite{frohlich}, when a charged particle is added to such a polar
insulating compound, the system responds by
screening the charge through the formation of a polaron, 
which is a combination of an electron (or hole)
plus the associated lattice distortion carrying the low frequency
polarization.
In these materials, dielectric screening is at the origin
of a possible strong electron-phonon coupling.  

Except for the
Anderson localization, which treats the effect of disorder on the
single particle properties, 
the theoretical 
scenarios for the MIT generally rely on collective mechanisms,
such as electron-lattice interactions (e.g. the
Peierls \cite{peierls} instability in one-dimensional metals) or
electron-electron  interactions. In the above mentioned Hubbard model, 
%
%
the insulating behavior comes from
the fact that putting two electrons (with opposite spins) on the same
electronic level of a single atom costs an energy $U$, which can be
larger than the kinetic energy gained in forming an electron band. 
Despite its simplicity, this model can be exactly
solved only in one dimension using the Bethe Antsatz  \cite{bethe}. In
higher dimensions, numerical calculations or sophisticated
approximations --- more or less
controlled --- are necessary to develop a physical insight
(see e.g.  \cite{Dagotto,georges,gebhard}). 
For example, the original proposal of Anderson  \cite{anderson},  
that a Resonating Valence Bond state should be the superconducting
ground-state of a two dimensional weakly doped
Hubbard model away from half filling, remains  unproved up to now.
Anyway, whatever the actual solution of
the model is (or will be), it cannot in itself justify the existence
of a finite critical  doping density to get the MIT (or a
superconducting transition),   as we discussed above. 

In what follows, we discuss in some detail two alternative scenarios,   
both relying on the long-range Coulomb interactions, that imply the
existence of a finite critical  doping density for the MIT.


\subsection {The Polarization Catastrophe: Herzfeld 1927  \cite{herzfeld}}

This is the first scenario which was proposed two years
before the Bloch theorem to understand why some elements of the
periodic table are metallic under normal conditions, whereas other
remain insulating. 
One starts from the element in its gaseous phase at
low density $N$  (let
us consider a metalloid such as Na  as an example). 
The Clausius-Mossotti (Lorenz-Lorentz) theory tells
us that the dielectric constant of such a gas satisfies: 
\begin{equation}
\frac {\epsilon -1}{\epsilon +2} = \frac {4 \pi}{3} N \alpha
\end{equation}
where $\alpha$ is the polarizability of a single isolated atom, due to  the
deformation of the electronic cloud around its nucleus. The original
argument of Herzfeld is based on  the fact that the dielectric
constant diverges (and changes sign) if the right hand side
of eq.(1) becomes larger than 1. A \textit{polarization catastrophe}
is thus expected  at $N_c = 3/4 \pi \alpha$, above which the system  becomes
metallic. This can be understood by taking a classical
model of an electron in its atomic orbital state 
around the nucleus, and giving rise to a dipole moment ${\bf p}=e {\bf
  r}$. The equation of motion is: 
\begin{equation}
 \ddot {\bf r} - \omega_0^2 {\bf r} = (e/m) {\bf E}_{loc}
\end{equation}
where $\hbar \omega_0$ is the electronic transition energy between two
atomic levels, related to the polarizability of the atom by
$\alpha = e^2/m \omega_0^2$. 
The local electric field on the right hand side 
contains a Lorentz field factor ${\bf E}_{loc}=E+4 \pi e N {\bf
  p}$ which accounts for the interactions with the other atoms in
the system. As a result, the frequency of the  
restoring force acting on the electron under study is softened as: 
\begin{equation}
\omega^2 = \omega_0^2 - \omega_P^2/3,
\end{equation} 
where $\omega_P^2=4 \pi Ne^2/m$ defines the plasma frequency. For
$N>N_{c_1} $, the restoring force on the electrons vanishes, 
which gives a criterion for the \textit{instability of the insulating phase}. 
For a review on recent applications of Herzfeld's criterion 
to understand the metallization of some elements, 
the reader is referred to the papers of P. P. Edwards 
 \cite{edwards1, edwards2}. 
The criterion can be recast in a different form \cite{edwards2}, which is
useful for comparison with the Mott transition (cf. next section). The
polarizability of a neutral atom can be  related to the
radius $a_H$ of the electronic orbital by $\alpha \approx (9/2) a_H^3$
 \cite{ashcroft}. In that case the critical density, related to $a_H$,
is: 
\begin{equation}
\label{herzfeldequ}
{N_{c_1}}^{1/3} a_H \approx 0.38
\end{equation}
For $N>N_{c_1}$ the insulating phase is unstable with repect to
metallization.

\subsection{The Mott Transition: 1961  \cite{mott}}

Much later,  a different mechanism for the MIT was proposed by Mott. 
He argued that in the metallic phase,
the Coulomb potential of the atoms would be screened by the mobile
electrons, and behave as $V(r) \approx (-e^2/r) \exp(-k_{TF}\ r)$,
$k_{TF}$ being the inverse Thomas-Fermi screening length 
 \cite{ashcroft}. The potential $V(r)$ does not allow for a localized
solution if $k_{TF}^{-1} >\approx  a _H$, where $a_H$ is the Bohr radius of
the electron in the unscreened potential $ \sim -e^2/r$. Consequently,
the critical density for the occurence of the insulating phase is: 
\begin{equation}
\label{mottequ}
N_{c_2}^{1/3} a_H  \approx 0.26
\end{equation} 
Since it takes into
account the screening of the Coulomb interaction in the metallic
phase, contrary to the Herzfeld citerion, which is a condition of
instability of the insulating phase, the Mott criterion is a condition
of \textit{instability of the metallic phase}.  As we shall see, this point is
of prime importance in our approach of the many-body treatment of the
polaron states.

\subsection{The role of polarons in the MIT mechanism}

We are left with two criteria describing respectively
an instability of the insulating and of the metallic phase.
There is no \textit{a priori} reason for the equality
$N_{c_1}=N_{c_2}$ to hold. If one looks to this problem in empty space, as we
discussed above and in view of the  estimates eq.(4) and (5),
$N_{c_1}>N_{c_2}$, which means that there is a region of densities
$N_{c_2}<N<N_{c_1}$, in which both states are stable (note that the above
discussion is for $T=0K$). The actual state is
determined by the one which has the lowest energy,
and it is generally the metallic state 
since the electrons delocalize in a Bloch band
(which is half filled in our simple problem) which greatly decreases
the energy with respect to localized states. In the intermediate
region, the insulating phase thus appears to be \textit{metastable}. 

In practice, 
the MIT under study  is observed by
chemically doping a given "host" material, whose 
dielectric properties must be taken into account. Let us
distinguish between two main classes of materials. 
The first class  consist of non-polar (or weakly polar) materials such as
conventional semiconductors (Si for example), for which the dielectric
constant is almost frequency independent,
$\epsilon_{host} (\omega) \approx \epsilon_{host}=constant$. In that
case, the situation is analogous  as in empty space, except that the Bohr
radius which enters in the different criteria must be modified as
$a_H^*=\epsilon_{host}a_H$. Most studies on the MIT in such compounds
confirm this scenario  \cite{edwards2}. The recent
discovery of superconductivity in doped diamond close to the MIT
density, could well revive the problem  \cite{diamond}, although
diamond in principle belongs to this first class of materials. 
 
The other class corresponds to polar materials, such as metal
oxide insulators (as the cuprates). In that case, complications arise from
the existence of two sources of dielectric screening. One is
the atomic polarization, responsible for the high
frequency dielectric constant $\epsilon_{\infty}$.
The second 
is due to
the displacement of the ions of the host material from their equilibrium
positions, whose relevant frequency scale is $ \omega_{LO}$, the
longitudinal optical frequency of the phonons. 
Such source of polarization acts  when the particles localize, 
and leads (together with the atomic
polarization) to the low
frequency dielectric constant $\epsilon_s$, which can be much
larger than $\epsilon_{\infty}$.
In a metallic state at high doping levels, 
the electronic plasma frequency 
will be changed into $\omega_P/\sqrt{\epsilon_{\infty}}$ and
not reduced by $\epsilon_s$. 
However, at low doping, the second kind of polarization acts to
localize each electron (or hole) wavefunction  in a bound state --- 
a polaron --- by creating a  potential-well which is Coulombic at
large distances.
In the strong coupling limit, the localization (polaron) radius
$R_P$ is essentially that of an electron in a potential $V(r)
\approx -e^2/ \tilde \epsilon r$, with an effective dielectric
constant $\tilde
\epsilon^{-1}=\epsilon_{\infty}^{-1}-\epsilon_s^{-1}$. 
Polarons are formed because the
gain in electronic energy due to the ionic distortion is  
always larger than the cost in elastic and localization energy
 \cite{frohlich}. 

The host material is thus able to induce  
the formation of polarons,
and these can play a central role in the
MIT. Applying the Mott criterion  \cite{queque2} to such bound states
gives:  
\begin{equation}
N_{c_2}^{1/3} \left( \tilde \epsilon / \epsilon_{\infty} \right) R_P \approx 
0.26, 
\end{equation}
which tells us that for $N<N_{c_2}$, some polarons must be formed out of 
the homogeneous electron gas.

The calculation of $N_{c_1}$ is much more difficult 
because polarons are charged particles, 
and their mutual Coulomb interactions 
in the insulating phase must be taken into account to obtain reliable
results. This task has been carried out through the study of the melting of a
Wigner Crystal of Polarons  (see next section), 
but the main conclusions apply more generally to any insulating 
polaronic state. Our main result
is the following:  when the electron-phonon
coupling is strong, there is no possibility of getting a liquid  state of
polarons with metallic properties. Instead,  a polarization catastrophe
occurs  at a density $N_{c_1}$, above which 
some of the polarons must dissociate. An
optical signature of this scenario has been derived: 
the peak in the optical conductivity due to the polaronic bound states
in the insulating phase, is shifted towards lower frequency as the density
increases. Secondly,  owing to the polaron-polaron interactions,
$N_{c_1}<N_{c_2}$, so that a range of densities exists for which
both a metallic state of free electrons and an insulating state of polarons
are unstable, as is sketched on Fig.1. In between, the system could
be electronically separated between a concentration $N_{c_1}$ of
localized polarons, and $N_{free}=N-N_{c_1}$ of free electrons (or
holes). 

\begin{figure}[htbp]
  \centering
  \resizebox{13cm}{!}{\includegraphics{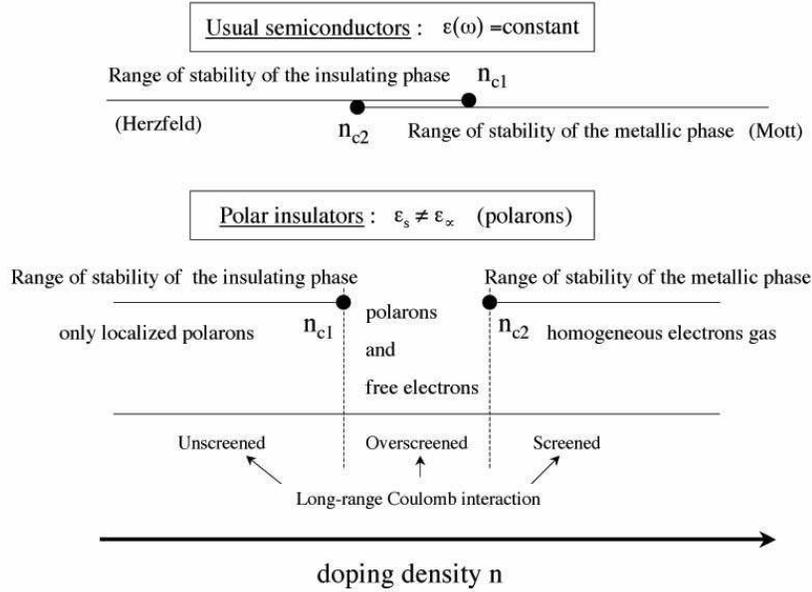}}
  \caption{A sketch of the scenario of the metal-insulator transition 
  in nonpolar and polar materials. In usual semiconductors with
  frequency independent dielectric constants, the Herzfeld
  instability (polarization catastrophe) takes place at higher
  densities than the Mott instability. For polar insulators, the
  interactions between polarons shift the Herzfeld instability to
  lower densities, leaving a range of densities where polarons
  could coexist with free electrons (or holes). In this region the
  Coulomb interactions between free electrons can be overscreened by
  the polaronic collective excitations (see text). 
 }
  \label{fig:pd}
\end{figure}

A system of crystallized polarons at density $N_{c_1}$
has a peculiar dielectric response, with a negative dielectric
constant down to zero frequency. That means that the Coulomb
interactions between free electrons can be overscreened by the
electrons localized in the polaronic states, and could lead
to a superconducting ground state, transforming the MIT 
into an insulator-superconductor transition. 
In fact, the outcome depends crucially  on the
behaviour of the counter-ionic charges (the doping ions). If
these are frozen in the host structure (the case of Sr in
La$_{2-x}$Sr$_x$CuO$_4$ for example), the scenario is viable. But if the
counter-ions are free to move, they also respond to the negative
dielectric constant of the electronic system, resulting in  
a true phase separation (as is
the case for O in La$_2$CuO$_{4+y}$, or for metal atoms in the
metal-ammonia solutions, cf. below). In the present
scenario, long range interactions between polarons are responsible
either for a superconducting instability, 
or for a macroscopic phase separation. 

\section{The Wigner Crystal of Polarons}

To systematically study the interactions between polarons in the
insulating phase, we have carefully examined the melting 
of a Wigner crystal of polarons as a
function of density. It was first
recognized in  \cite{queque1} that the ground-state of polarons at low
densities should be a Wigner crystal  \cite{wigner}. Detailed
studies have been carried out on this problem in 
references \cite{tout1,tout2,tout3,rastelli},  
to which the reader is referred for more details. 

Wigner crystallization
occurs at low densities because 
the average electron-electron interaction energy 
(proportional to $1/r_s$, where $r_s$ is the mean distance between
electrons at densities $n \sim r_s^{-3}$) is much larger than the kinetic
energy (proportional to $1/r_s^2$), so that the ground state is 
crystallized 
in order to  minimize the potential energy.
In a host polar material, a Wigner crystal of
electrons is transformed into a Wigner crystal of polarons, which has
two competing effects: 1) the Coulomb interactions between polarons are
reduced as $\sim 1/\epsilon_s r$, which tends to destabilize the
crystallized state; but 2) the effective mass of the carriers
(polarons) is increased because each electron carries its own polarization
cloud, and this tends to stabilize the crystallized state. In 
a highly polarizable material ($\epsilon_s \gg \epsilon_\infty$), the
balance will essentially depend on the strength of the
electron-phonon coupling. In the Fr\"ohlich model, it is defined as
$\alpha = (m^*/2 \hbar^3 \omega_{LO})^{1/2}e^2/\tilde \epsilon$, $m^*$
being the band mass of a free electron \cite{reviewdevreese}. 
The strong
coupling regime, where the polaron 
behaves essentially  as a Coulombic bound state, is attained above
$\alpha \approx 6-7$, which is not common in real
materials. The cuprates seem to be an exception for
two main reasons: 1) the effective electronic band mass $m^*$ is
already high owing to the short range correlations, and can reach several 
units of the bare electron mass ($m_e$): $m^* \sim 2-4 m_e$  \cite{georges}; 2)
coherent electron motion is constrained in two-dimensional CuO$_2$ layers,
whereas the Coulomb interactions (and polarization  \cite{chen})
remain three-dimensional. As demonstrated by Devreese
 \cite{devreesepeter}, this shifts the strong electron-phonon coupling regime 
down to $\alpha \approx 3$. Due to the combination of these two aspects of the
problem, 
when excess holes (or electrons) are added to the parent insulating
cuprates, the formation of strong coupling Fr\"ohlich polarons cannot
be \textit{a priori} avoided.

\begin{figure}[htbp]
  \centering
  \resizebox{6.5cm}{!}{\includegraphics{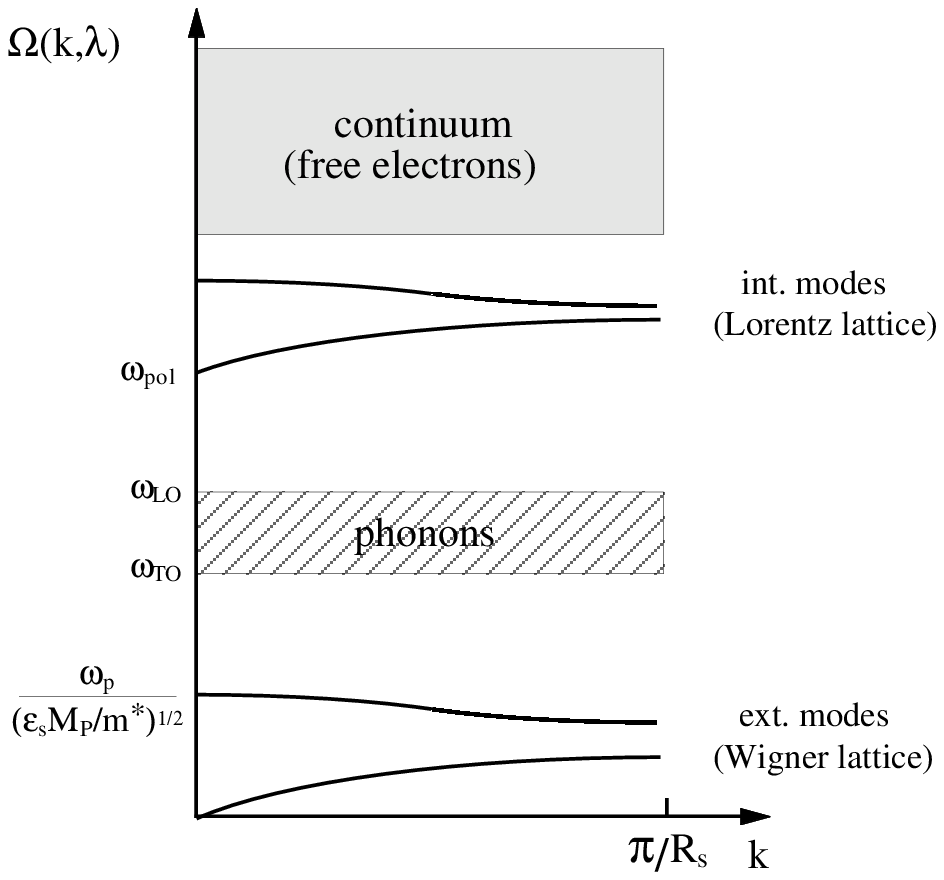}}
  \resizebox{6.5cm}{!}{\includegraphics{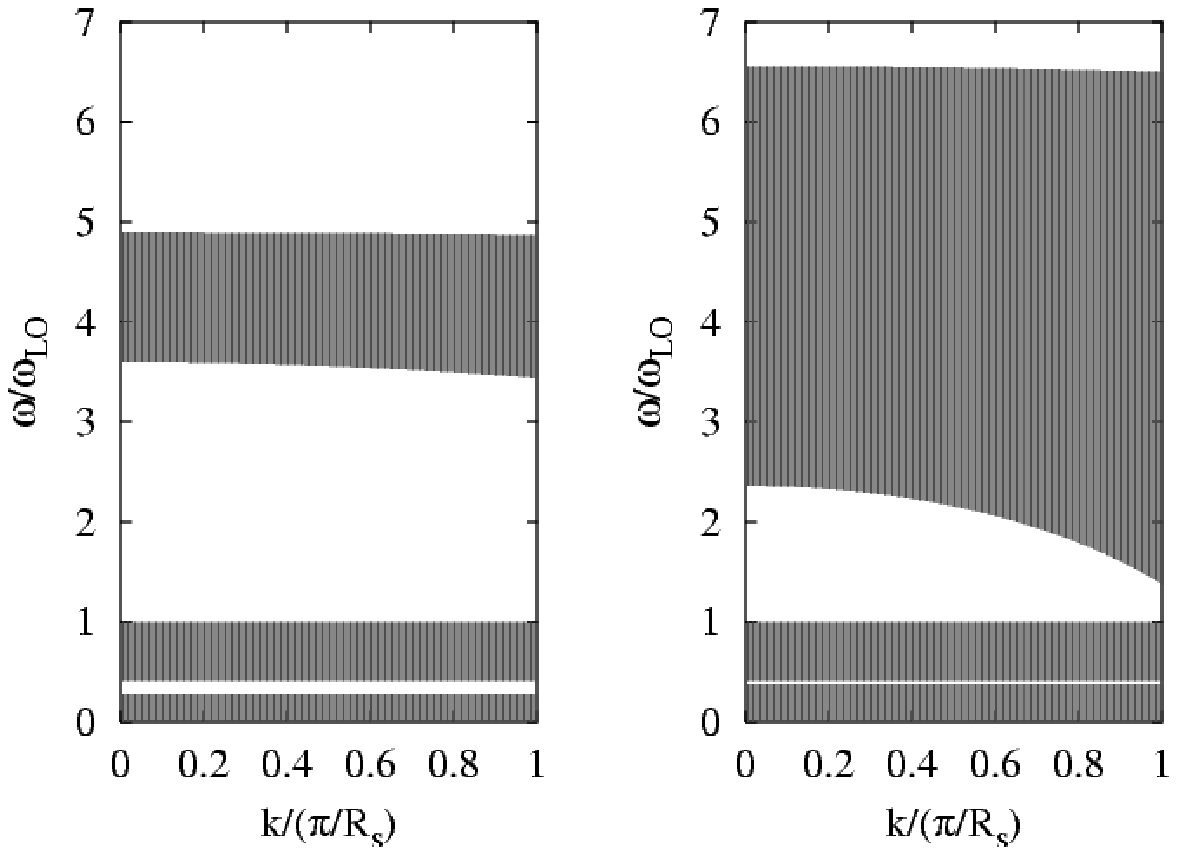}}
  \caption{Left panel: typical behaviour of the 
    collective excitations of a Wigner crystal of polarons 
    in the strong electron-phonon
    coupling regime. ``int.modes'' correspond to the vibrations 
    of the  electrons localized within their potential-wells (due to both
    electron-electron and electron-phonon interactions), whereas 
    ``ext.modes'' correspond to the low frequency vibrations of the
    polarons involving ionic displacements (see ref. [21]
    for details). $\omega_{pol}$ is the transverse collective mode observed in
    the optical absorption, which softens as the density increases up
    to the critical density for the polarization catastrophe, at which
     $\omega_{pol} = \omega_{LO}$. $M_p$ is the polaron mass. 
Right panel: Regions (in grey) in the $(k,\omega)$ plane where the dielectric
constant of the Wigner crystal of polarons is negative 
(giving rise to overscreening) for
$n=5 \cdot 10^{19}cm^{-3}$ and $n = 1.7\cdot  10^{20} cm^{-3}$. 
Such values  correspond to the parameters 
$\epsilon_s=30$, $\epsilon_\infty=5$,and $m^*=2m_e$ 
in the strong-electron phonon coupling regime, from ref. [21].}
  \label{fig:diel}
\end{figure}

Many theoretical difficulties
already arise on treating just one or two polarons.
The reason is that the problem of one single polaron is
already a many-body quantum problem which cannot be solved exactly. 
Feynman  \cite{feynman} provided the best solution through the
use of path-integrals, replacing this many-body problem by
a two body problem. In our studies of the polaron Wigner crystal, we
generalized the approach of Feynman to the many-polaron system,
taking advantage of the fact 
that  the
exchange between electrons can be neglected
in the crystallized phase at low densities.
Our basic hypothesis,
that two polarons repel with long-range Coulomb interaction, is
always fulfilled  provided that the dielectric constant satisfies
$\epsilon_{\infty}/\epsilon_s >0.1$  \cite{gitterman,
  devreesebip} (although other ad-hoc models
have been proposed to enforce the formation of bipolaronic bound states 
in the context of the cuprates).

Based on
Feyman's treatment of dielectric polarons, we have shown that the 
characteristic collective  frequencies in
the polaron Wigner crystal have the behaviour shown in Fig.2. A
polarization catastrophe was shown to occur upon increasing the density, when 
the condition
$\omega_{pol} \rightarrow \omega_{LO}$ is attained 
(the corresponding critical density $N_{c_1} \approx 5.10^{20}/cm^3$ obtained 
with the microscopic parameters of the cuprates is in good agreement with
experiment). $\omega_{pol}$ is the transverse optical  collective mode of the
Wigner crystal of polarons, and identifies the location of an absorption peak in the optical 
conductivity. A
simplified version for $\omega_{pol}$ is recovered in the limit
$\omega_{LO} \rightarrow 0$, which reproduces the original result obtained by
Bagchi  for a Lorentz lattice of dipoles  \cite{bagchi}: 
\begin{equation}
\omega_{pol}^2 = \omega_0^2- \frac {\omega_P^2} {3 \epsilon_{\infty}},
\end{equation}
where $\omega_0$ is the frequency of the electron localized
in its polaron potential-well.

As was mentioned above, the dielectric constant becomes
negative in large regions of $(k,\omega)$ 
as one reaches the critical density. This is illustrated in Fig.2 
(right panel), 
where the sign of the dielectric constant is shown for two
different densities. This result shows that, beyond the dielectric
catastrophe, free electrons (or holes) can be paired by the remaining
electrons localized in the polaronic states. 
\footnote{Note that the polarization catastrophe
is a general phenomenon in the case of neutral dipoles (cf. section 1.1).
For crystallized charged particles, on the other hand,
the softening of the
peak of conductivity can only occur if $\epsilon_s \neq
\epsilon_{\infty}$, i.e. owing to the polaron formation. The reason is
that if $\epsilon_s = \epsilon_{\infty}$, the collective frequencies in 
the crystallized state do not vary as the density is increased:
there is no soft mode in this case.} 

\section{Relevance of the polarization catastrophe scenario in real compounds}

The scenario described in the preceding section, which relies on 
very general and simple hypotheses, can in principle be observed in
the insulating (or poorly metallic) phases of any strongly
polarizable material, i.e.  as soon as the long range Coulomb interactions
are not screened. However, other ingredients are often present in the compounds
of interest, that can compete with the effects evidenced above, making
their clear identification difficult.

There are at least two classes of compounds where there are indeed
indications of the relevance of the polarization catastrophe scenario:
the superconducting cuprate materials and the metal-ammonia
solutions (MAS). The former are insulating  solids with a layered
crystal structure, that undergo a superconducting instability above a
certain critical doping level of the order of $5-10\%$. 
The latter are liquid  solutions that exhibit phase separation 
and become metallic above a given critical concentration of metal
ions,  around $3-8\%$. 
Beyond the complexities specific to each class
(let us mention again the ubiquitous antiferromagnetic correlations in the
cuprates, and the 
interplay with the classical dynamics of the complex fluid in the MAS),
both systems share the same two basic ingredients of the theory: 
they are strongly polarizable, and have unscreened interactions 
at low doping levels, due to the absence of mobile charges. 

\subsection{Cuprates}

\paragraph{Polaron softening}
The identification of the softening of the polaronic collective mode 
--- the clearest precursor
to the polarization catastrophe --- requires systematic doping dependent
measurements of the optical conductivity.  
Such studies have been performed in electron doped
Nd$_{2-x}$Ce$_x$CuO$_4$  \cite{Lupi99}, hole doped La$_{2-x}$Sr$_x$CuO$_4$
 \cite{Lucarelli03} and YBa$_2$Cu$_3$O$_y$ \cite{Lee05}, which
invariably exhibit an absorption peak around $\sim 0.15 eV$ that
progressively softens and hits the frequency range of phonon
excitations in correspondence with the superconducting instability
(there is also a stronger broad  peak at $\sim 0.5eV$, which  softens
in a parallel way but does not seem to undergo any drastic change at $x_c$).
The value $\sim 0.15 eV$ is compatible with the optical absorption of
dielectric polarons in such materials.

\paragraph{Charge modulations}

Recently, charge modulations of square symmetry and 
with a period of $\sim 4$ Cu-Cu lattice parameters,
have been observed by scanning tunneling microscopy
at the surface of Ca$_{2-x}$Na$_x$CuO$_2$Cl$_2$  \cite{Hanaguri} 
and Bi$_2$Sr$_2$CaCu$_2$O$_{8+\delta}$
 \cite{MCElroy} and interpreted as an unusual charge ordered
state, possibly related to Wigner crystallization. Although it is not
clear experimentally if this is a genuine charge ordering of carriers,
or if the observed modulations involve pairs of carriers, a
calculation based on a phenomenological Lorentz model \cite{EPJBIII} 
shows that Wigner crystallization of holes is compatible with the
observed periodicity, provided that an additional source of
carrier localization is included. The energy scale of such additional
mechanism $\sim 0.15 eV$, deduced from the model, agrees with what
is measured in optical experiments and could well be of polaronic
origin.  Furthermore, the  square
symmetry of the observed charge ordering  follows naturally from the present scenario if one accounts 
for the \textit{isotropic} long-range
repulsion between the holes, i.e. including the unscreened Coulomb interactions
between different layers.

\bigskip

The scenario emerging from several experiments
in the cuprates (including optical and photoemission experiments)
points to the coexistence of  localized carriers of polaronic character with 
free-electron like carriers. It is still unclear if these different
``fluids'' are spatially separated in ordered or disordered patterns,
if they live at different 
energy scales or in different parts of the Brillouin zone.

\subsection{Metal-ammonia solutions}

\begin{figure}[htbp]
  \centering
  \resizebox{6.5cm}{!}{\includegraphics{fig3_prl99.epsf}}
  \resizebox{6.5cm}{!}{\includegraphics{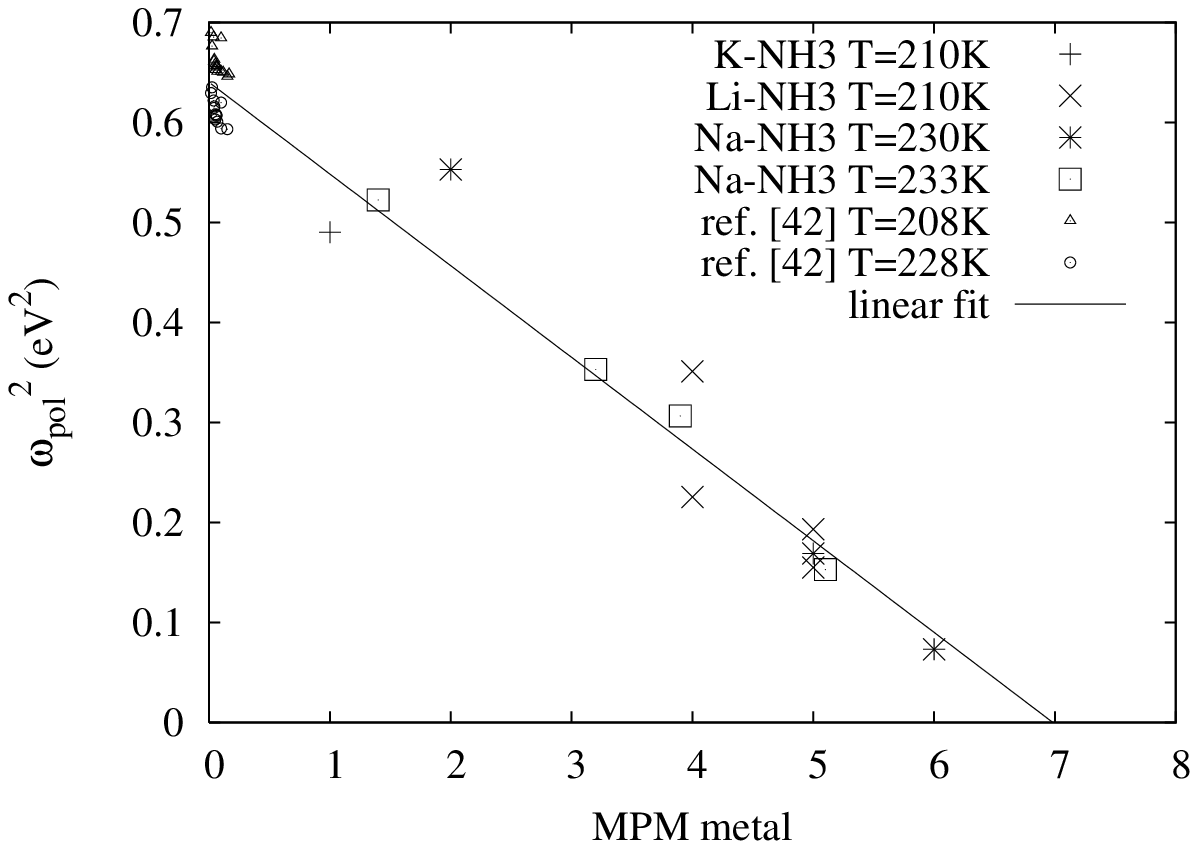}}
  \caption{Softening of the absorption peak in Nd$_{2-x}$Ce$_x$CuO$_4$, vs. 
  carrier concentration (left panel, courtesy of
    S. Lupi, from ref. [30]. IST indicates the insulator-to-superconductor 
    transition) and in   
    metal-ammonia  solutions vs. concentration of metal ions, 
    from reference [37] (right panel).  Data were collected by fitting the
    absorption and permittivity data in refs [41,42] with a Lorentz
    model.}
  \label{fig:ammonia}
\end{figure}

When an excess electron is introduced in liquid ammonia, a bound state
is formed which involves the long-range polarization field coming from
the orientational polarizability of the ammonia molecules. Such bound state is
analogous to a polaron in a ionic dielectric \cite{Jortner}, and
manifests through a broad optical absorption peak at $\sim 0.8 eV$.
Upon increasing the concentration,
the frequency of such absorption peak 
exhibits a clear softening, whose square  
follows a linear trend as expected from eq. (7).
It extrapolates to $0$  roughly at the boundary
of the phase separation region \cite{unpub} (see figure 3, right panel), 
indicating the possible relevance of
the polarization catastrophe scenario in the mechanism of the
metal-nonmetal transition.

An analogy between the microscopic mechanisms underlying the phase
diagrams of the cuprates and the metal-ammonia solutions has
been proposed in reference  \cite{analogy}.

\subsection{Other examples}

The concept of Wigner crystallization of  polarons has been applied
recently to other classes of compounds. In ref.  \cite{Devreese}, a
lattice of ripplopolarons has been shown to arise at the surface of 
a multielectron bubble in liquid Helium.
In ref.  \cite{castroneto}, 
the crystal of magnetic polarons arising as the low density solution of
the double-exchange model has been discussed in the framework of 
the magnetic hexaborides EuB$_6$.

\acknowledgments

\end{document}